\documentclass[conference]{IEEEtran}

\renewcommand\IEEEkeywordsname{Index Terms}
\usepackage{graphicx}
\usepackage{subcaption}
\usepackage{tikz,xparse}

\usetikzlibrary{patterns}

\usetikzlibrary{dsp,chains}
\usetikzlibrary{matrix}
\usepackage{mathptmx}
\usepackage{verbatim}
\usepackage{calc}% http://ctan.org/pkg/calc
\usepackage{ifthen}
\usepackage{xifthen}
\usepackage{cancel}
\usepackage{bm}
\usepackage{verbatim}
\usepackage{multirow}
\usepackage{cite}
\usepackage[hyphens]{url}
\usepackage[nolist]{acronym} 
\usepackage{pgfplots}
\usetikzlibrary{arrows,shapes,graphs,graphs.standard,quotes,arrows.meta,decorations.markings}
\usetikzlibrary{matrix}
\pgfplotsset{compat=newest}
\usepackage[bookmarks=false]{hyperref}
\usepackage{units}
\usepackage{amsmath, amsbsy, amssymb, latexsym }
\hypersetup{bookmarksdepth=-2}
\usepackage{comment}
\usepackage[utf8]{inputenc}
\usepackage{xcolor}
\usepackage{enumitem}
\usepackage{algorithm}
\usepackage{algpseudocode}
\usetikzlibrary{spy}

\usepackage{marginnote}

\tikzset{>=latex}

\algnewcommand\algorithmicforeach{\textbf{for each}}
\algdef{S}[FOR]{ForEach}[1]{\algorithmicforeach\ #1\ \algorithmicdo}

\algnewcommand\algorithmicswitch{\textbf{switch}}
\algnewcommand\algorithmiccase{\textbf{case}}
\algnewcommand\algorithmicassert{\texttt{assert}}
\algnewcommand\Assert[1]{\State \algorithmicassert(#1)}%
\algdef{SE}[SWITCH]{Switch}{EndSwitch}[1]{\algorithmicswitch\ #1\ \algorithmicdo}{\algorithmicend\ \algorithmicswitch}%
\algdef{SE}[CASE]{Case}{EndCase}[1]{\algorithmiccase\ #1}{\algorithmicend\ \algorithmiccase}%
\algtext*{EndCase}%

\definecolor{mittelblau}{RGB}{0, 126, 198}
\definecolor{violettblau}{cmyk}{0.9, 0.6, 0, 0}
\definecolor{rot}{RGB}{238, 28 35}
\definecolor{apfelgruen}{RGB}{140, 198, 62}
\definecolor{gelb}{RGB}{255, 229, 0}
\definecolor{orange}{RGB}{244, 111, 33}
\definecolor{pink}{RGB}{237, 0, 140}
\definecolor{lila}{RGB}{128, 10, 145}
\definecolor{hellgrau}{RGB}{224, 224, 224}
\definecolor{mittelgrau}{RGB}{128, 128, 128}
\definecolor{dunkelgrau}{RGB}{80,80,80}
\definecolor{anthrazit}{RGB}{19, 31, 31}
\definecolor{darkgreen}{RGB}{34,139,34}
\definecolor{aqua}{RGB}{0, 255, 255}
\colorlet{Mycolor1}{green!10!orange!90!}
\tikzset{
       vnd/.style={
        shape=circle,
        fill=black,
        draw,
        inner sep=0pt,
        minimum size=0.2cm},
        cnd/.style={
        shape=rectangle,
        fill=white,
        draw,
        minimum width=0.05mm,
        minimum height = 0.05mm}, 
         vndR/.style={
        shape=circle,
        fill=red,
        draw,
        inner sep=0pt,
        minimum size=0.2cm},
        cndR/.style={
        shape=rectangle,
        fill=white,
        draw=red,
        minimum width=0.05mm,
        minimum height = 0.05mm}
}

\DeclareMathOperator*\bigboxplus{\ensuremath{\boxplus}}

\newcommand{\new}[1]{{\color{black}#1}}

\IEEEoverridecommandlockouts

\renewcommand{\vec}[1]{\mathbf{#1}}

\newcommand{\bv}{\vec{b}}

\newcommand{\uv}{\vec{u}}

\newcommand{\xv}{\vec{x}}
\newcommand{\yv}{\vec{y}}
\newcommand{\zv}{\vec{z}}
\newcommand{\zerov}{\vec{0}}

\newcommand{\Am}{\vec{A}}

\newcommand{\Gm}{\vec{G}}
\newcommand{\Hm}{\vec{H}}

\newcommand{\FF}{\mathbb{F}}

\newcommand{\LB}{\left(}
\newcommand{\RB}{\right)}
\newcommand{\LP}{\left\{}
\newcommand{\RP}{\right\}}

\newcommand{\sgn}{\mathop{\mathrm{sgn}}}

\newcommand{\argmin}{\mathop{\mathrm{argmin}}}
\newcommand{\argmax}{\mathop{\mathrm{argmax}}}

\begin{document}
	
\begin{NoHyper}
\title{Iterative Reed--Muller Decoding}

\author{\IEEEauthorblockN{Marvin Geiselhart$^*$, Ahmed Elkelesh$^*$, Moustafa Ebada$^*$, Sebastian Cammerer$^\dagger$ and Stephan ten Brink$^*$}  \thanks{An extended version of this paper is available \cite{rm_automorphism_ensemble_decoding}.}
	\IEEEauthorblockA{
		$^*$Institute of Telecommunications, Pfaffenwaldring 47, University of  Stuttgart, 70569 Stuttgart, Germany 
		\\\{geiselhart,elkelesh,ebada,tenbrink\}@inue.uni-stuttgart.de\\
		$^\dagger$NVIDIA, Fasanenstraße 81, 10623 Berlin, Germany\\
		scammerer@nvidia.com
	}
\vspace{-.6cm}
}

\maketitle

\begin{acronym}
\acro{ECC}{error-correcting code}
\acro{HDD}{hard decision decoding}
\acro{SDD}{soft decision decoding}
\acro{ML}{maximum likelihood}
\acro{GPU}{graphical processing unit}
\acro{BP}{belief propagation}
\acro{BPL}{belief propagation list}
\acro{LDPC}{low-density parity-check}
\acro{HDPC}{high density parity check}
\acro{BER}{bit error rate}
\acro{SNR}{signal-to-noise-ratio}
\acro{BPSK}{binary phase shift keying}
\acro{AWGN}{additive white Gaussian noise}
\acro{MSE}{mean squared error}
\acro{LLR}{Log-likelihood ratio}
\acro{LUT}{look-up table}
\acro{MAP}{maximum a posteriori}
\acro{NE}{normalized error}
\acro{BLER}{block error rate}
\acro{PE}{processing elements}
\acro{SCL}{successive cancellation list}
\acro{SC}{successive cancellation}
\acro{SCAN}{soft cancellation}
\acro{BI-DMC}{Binary Input Discrete Memoryless Channel}
\acro{CRC}{cyclic redundancy check}
\acro{CA-SCL}{CRC-aided successive cancellation list}
\acro{BEC}{Binary Erasure Channel}
\acro{BSC}{Binary Symmetric Channel}
\acro{BCH}{Bose-Chaudhuri-Hocquenghem}
\acro{RM}{Reed--Muller}
\acro{RS}{Reed-Solomon}
\acro{SISO}{soft-in/soft-out}
\acro{PSCL}{partitioned successive cancellation list}
\acro{3GPP}{3rd Generation Partnership Project }
\acro{eMBB}{enhanced Mobile Broadband}
\acro{PCC}{parity-check concatenated}
\acro{CA-polar codes}{CRC-aided polar codes}
\acro{CN}{check node}
\acro{VN}{variable node}
\acro{PC}{parity-check}
\acro{GenAlg}{Genetic Algorithm}
\acro{AI}{Artificial Intelligence}
\acro{MC}{Monte Carlo}
\acro{CSI}{Channel State Information}
\acro{PSCL}{partitioned successive cancellation list}
\acro{OSD}{ordered statistic decoding}
\acro{MWPC-BP}{minimum-weight parity-check BP}
\acro{FFG}{Forney-style factor graph}
\acro{MBBP}{multiple-bases belief propagation}
\acro{NBP}{neural belief propagation}
\acro{URLLC}{ultra-reliable low-latency communications}
\acro{DMC}{discrete memoryless channel}
\acro{MSB}{most significant bit}
\acro{LSB}{least significant bit}
\acro{RPA}{recursive projection-aggregation}
\acro{SGD}{stochastic gradient descent}
\end{acronym}

\begin{abstract}
Reed--Muller (RM) codes are known for their good \ac{ML} performance in the short block-length regime. Despite being one of the oldest classes of channel codes, finding a low complexity soft-input decoding scheme is still an open problem. In this work, we present a belief propagation (BP) decoding architecture for RM codes based on their rich automorphism group. The decoding algorithm can be seen as a generalization of multiple-bases belief propagation (MBBP) using polar BP as constituent decoders. We provide extensive error-rate performance simulations and compare our results to existing decoding schemes. We report a near-ML performance for the RM(3,7)-code (e.g., $0.05$ dB away from the ML bound at BLER of $10^{-4}$) at a competitive computational cost.
To the best of our knowledge, our proposed decoder achieves the best performance of all iterative RM decoders presented thus far.
\end{abstract}
\acresetall

%\begin{IEEEkeywords}
%Reed--Muller Codes, Polar Codes, Code Automorphisms, Successive Cancellation Decoding, Belief Propagation Decoding, List Decoding, Ensemble Decoding.
%\end{IEEEkeywords}

\acresetall
%\vspace{-0.2cm}
\section{Introduction}
The current trend of \ac{URLLC} applications has urged the need for efficient short length coding schemes in combination with the availability of efficient decoders.
Besides many other coding schemes, this has lead to the revival of one of the oldest error-correcting codes, namely \ac{RM} codes \cite{Muller1954,Reed1954}
-- potentially also due to some existent similarities between \ac{RM} codes and the newly developed family of polar codes \cite{ArikanMain, StolteRekursivPlotkin}. 
On the one hand, \ac{RM} codes, as an example of algebraic codes, are known to be capacity-achieving over the \ac{BEC} for a given rate \cite{Abbe_RM_BEC,RM_Capacity_BEC}. Moreover, and practically even more relevant, they enjoy an impressive error-rate performance under \ac{ML} decoding, which also holds in the short length regime.
To this extent, several decoding algorithms have been developed in the course of \ac{RM} decoding. On the other hand, to the best of our knowledge, there is still a lack of practical decoders that are characterized by near-ML performance \emph{and} feasible decoding complexity/latency.

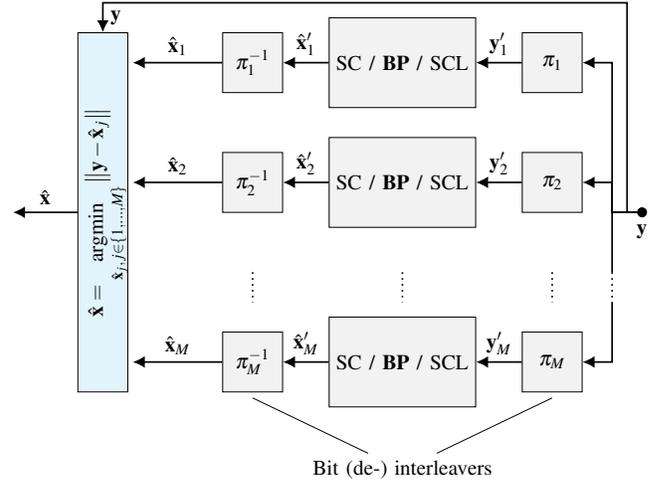
\begin{figure} [t]
	\centering
	\resizebox{0.975\columnwidth}{!}{\begin{tikzpicture}
\tikzset{
edge/.style = {thick,black},
mydiamond/.style={draw, diamond, aspect=2.7,text width=1.75cm, inner sep=0pt,  fill=white!90!red},
BPrectangle/.style={rectangle, draw, minimum size=1.5cm, fill=white!90!gray},
intrect/.style={rectangle, draw, minimum size=1cm, fill=white!90!gray}
}

\tikzstyle{conn} = [-{Latex[length=2mm,width=2mm]}];

\node[draw,shape=circle, fill=black, inner sep=0pt,minimum size=0.15cm, label=below:{$\mathbf{y}$}] (Lch) at (2, 0.5) {};

\draw [edge, dotted] (-2,-0.5)--(-2,-1);
\draw [edge] (Lch)--(1.5,0.5)--(1.5,-0.5);
\draw [edge, dotted] (1.5,-0.5)--(1.5,-1);

\draw [edge, black,-, dotted] (-4.5,-0.5)--(-4.5,-1);

\draw [edge, black,-, dotted] (.5,-0.5)--(.5,-1);

\node[BPrectangle] (BP1) at (-2, 3) {SC / \textbf{BP} / SCL};
\node[intrect] (int1) at (.5, 3) {$ \pi_1 $};
\node[intrect] (deint1) at (-4.5, 3) {$ \pi_1^{-1} $};
\draw [edge,conn] (Lch)--(1.5,0.5)--(1.5,3)--(int1);

\node[BPrectangle] (BP2) at (-2, 1) {SC / \textbf{BP} / SCL};
\node[intrect] (int2) at (.5, 1) {$ \pi_2 $};
\node[intrect] (deint2) at (-4.5, 1) {$ \pi_2^{-1} $};
\draw [edge,conn] (Lch)--(1.5,0.5)--(1.5,1)--(int2);

\node[BPrectangle] (BP3) at (-2, -2) {SC / \textbf{BP} / SCL};
\node[intrect] (int3) at (.5, -2) {$ \pi_M $};
\node[intrect] (deint3) at (-4.5, -2) {$ \pi_M^{-1} $};
\draw [edge,conn] (1.5,-1)--(1.5,-2)--(int3);

\draw [edge,conn](int1) to node[above] {$\mathbf{y}'_1$} (BP1);
\draw [edge,conn](int2) to node[above] {$\mathbf{y}'_2$} (BP2);
\draw [edge,conn](int3) to node[above] {$\mathbf{y}'_M$} (BP3);

\draw [edge,conn](BP1) to node[above] {$\hat{\mathbf{x}}'_1$} (deint1);
\draw [edge,conn](BP2) to node[above] {$\hat{\mathbf{x}}'_2$} (deint2);
\draw [edge,conn](BP3) to node[above] {$\hat{\mathbf{x}}'_M$} (deint3);

\draw [edge,conn](deint1) to node[above] {$\hat{\mathbf{x}}_1$} (-6.5,3);
\draw [edge,conn](deint2) to node[above] {$\hat{\mathbf{x}}_2$} (-6.5,1);
\draw [edge,conn](deint3) to node[above] {$\hat{\mathbf{x}}_M$} (-6.5,-2);

\node[] (int) at (-2,-3.8) {Bit (de-) interleavers};
\draw [black,-, ] (int)--(.5,-2.6);
\draw [black,-] (int)--(-4.5,-2.6);

\node[rectangle, draw, minimum width=6cm, minimum height=0.3cm, text height=0.3cm, text depth=0.3cm,text centered,rotate=90, fill=white!90!cyan] (decide) at (-7, 0.5) {$\hat{\mathbf{x}}=\underset{\hat{\mathbf{x}}_{j},j\in\left\{ 1,\dots,M\right\} }{\mathrm{argmin}}\left\Vert \mathbf{y}-\hat{\mathbf{x}}_{j}\right\Vert $};

\draw [edge,conn] (Lch)--(1.75,0.5)--(1.75,4)--(-7,4)to node[right] {$\mathbf{y}$}(decide);

\draw [edge,conn](decide) to node[above] {$\hat{\mathbf{x}}$} (-8.5,0.5);

\end{tikzpicture}}
	\caption{\footnotesize Block diagram of automorphism ensemble decoding. We focus on the case where $ M $ constituent \ac{BP} decoders are used.
		The usage of \ac{SC} and \ac{SCL} as constituent decoders is presented in the extended version of this paper \cite{rm_automorphism_ensemble_decoding}.}	
	\label{fig:Block_Diag}
	\vspace{-.4cm}
\end{figure}

\ac{RM} decoders can be grouped into two main categories, iterative and non-iterative decoders which we will shortly revisit in the following.
In the literature, the best known decoder for RM codes over an \ac{AWGN} channel is Dumer’s recursive list decoding algorithm \cite{permuteRM}, \new{which is now known under the name \ac{SCL} decoding \cite{talvardyList}, and a variant using permutations.}
Recently, a \ac{RPA} decoding algorithm for RM codes was proposed in \cite{RPA_Abbe}, which can be viewed as a weighted BP decoder over a redundant factor graph \cite{WBP_RPA_Pfister_ISIT20}, making use of the symmetry of the RM codes (i.e., its large automorphism group).
RM codes under RPA decoding were shown to outperform the error-rate performance of CRC-aided polar codes under \ac{SCL} decoding.
A more general usage of the rich automorphism group of RM codes to aid the decoding process is reported in \cite{StolteRekursivPlotkin}, along with the decoding of RM codes using a redundant parity-check matrix proposed earlier in \cite{Bossert_RM}.
In this work, for the sake of comparison, we consider the following iterative decoders: multiple-bases belief propagation\acused{MBBP} (\ac{MBBP}) decoding \cite{Huber}, minimum weight parity-check belief propagation\acused{MWPC-BP} (\ac{MWPC-BP}) decoding \cite{PfisterMWPC}, neural belief propagation\acused{NBP} (\ac{NBP}) decoding \cite{NachmaniNeuralBP}, pruned neural belief propagation (pruned-\ac{NBP}) decoding \cite{buchberger2020pruning}. 
The inherent parallel nature of iterative decoders allows high throughput implementations. 
Besides, their \ac{SISO} nature makes them suitable for iterative detection and decoding.

In this paper, we propose a new iterative decoding scheme, extending and generalizing some of the previously mentioned decoding algorithms. 
To the best of our knowledge, our proposed decoding algorithm achieves the best practical iterative decoding performance of the RM(3,7)-code presented thus far ($0.05$ dB away from the ML bound at BLER of $10^{-4}$).
Fig.~\ref{fig:Block_Diag} shows an abstract view of our proposed decoding algorithm.

\section{Preliminaries}\label{sec:preliminaries}
\subsection{Reed--Muller Codes}
We interpret each message of an RM($ r $,$ m $) code as a multi-linear polynomial $ u(\zv) $ in $ m $ binary variables $ z_j $ (with $j\in\{0,\cdots,m-1\}$) and maximum degree $ r $, over the finite field $ \FF_2 $. To obtain the codeword $ \xv $, the message polynomial is evaluated at all points in the space $ \FF_2^m $, resulting in $ N = 2^m $ codeword bits \cite{Muller1954, Reed1954}.

\subsection{Automorphism Group}
The \emph{automorphism group} (or \emph{permutation group}) $ \operatorname{Aut}(\mathcal{C}) $ of a code $ \mathcal{C} $ is the set of permutations $ \pi $ of the codeword bit indices that map $ \mathcal{C} $ onto itself, i.e. 
\begin{equation}
	\pi(\xv) \in \mathcal{C} \quad \forall \xv \in \mathcal{C} \quad \forall \pi \in \operatorname{Aut}(\mathcal{C}),
\end{equation}
where $ \pi(\xv) $ results in the vector $ \xv' $ with $ x_i' = x_{\pi(i)} $. In other words, every codeword is mapped to another (\emph{not} necessarily different) codeword of the same code.
The automorphism group of \ac{RM} codes is well known as the \emph{general affine group} GA($ m $) over the field $ \FF_2 $\cite{macwilliams77}.\footnote{In this paper, we only consider the field $ \FF_2 $ and hence, we omit the size of the field in the notation, i.e., we write GA($ m $) instead of GA($ m,2 $).} GA($ m $) is the group of all affine bijections over $ \FF_2^m $, i.e., pairs $ (\Am, \bv) $ defining the mapping $ \zv' = \Am \zv + \bv $, with a non-singular matrix $ \Am \in \FF_2^{m\times m}$ and an arbitrary vector $ \bv \in \FF_2^{m\times 1} $. The vectors $ \zv$, $\zv' \in \FF_2^{m\times 1} $ are the binary representations of the code bit positions $ i $ and $ \pi(i) $, respectively, i.e., $ i = \sum_{j=0}^{m-1}{z_j \cdot 2^j} $. An important subgroup of GA($ m $) is the set of stage-shuffle permutations $ \Pi(m) $, corresponding to the special case where $ \Am $ is a permutation matrix and $ \bv=\zerov $.

\subsection{Iterative RM Decoding}
In this section, we briefly revise the different iterative decoding techniques which can be used for \ac{RM} codes. 

\subsubsection{Na\"ive Belief Propagation Decoding}
\ac{BP} decoding can be performed over the Tanner graph of a code's parity-check matrix. However, for high-density parity-check codes like \ac{RM} codes, the performance of this decoder is usually poor due to the numerous short cycles in the graph.

\subsubsection{Belief Propagation Decoding over Forney-style Factor Graph}\label{sec:ffg}
Rather than on a Tanner graph, \ac{BP} decoding can \new{also be} performed over a \ac{FFG}, constructed from check and variable nodes of degree three \cite{Forney}. Fig.~\ref{fig:rm_ffg} shows the \ac{FFG} of the RM(1,3)-code.

As the frozen nodes always contribute the same \acp{LLR} (i.e., a priori \acp{LLR} of $ +\infty $) to the equations, the \ac{FFG} can be reduced as shown on the right in Fig.~\ref{fig:rm_ffg}, by removing edges of constant value. This potentially reduces the number of performed arithmetic operations per iteration while preserving the same performance \cite{Forney}. In this work, whenever the BP decoding is used over the \ac{FFG}, we use the reduced version.

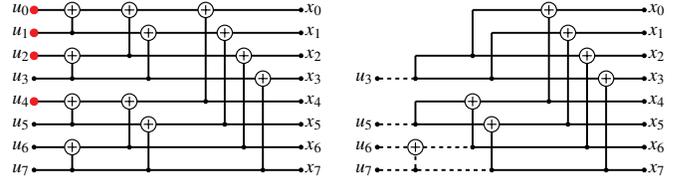
\begin{figure}[t]
	\centering\resizebox{\columnwidth}{!}{\begin{tikzpicture}[yscale=.6]
\tikzstyle{every node}=[font=\Large]
\tikzstyle{frozennode} = [dspnodefull,minimum size=2mm,rot];
\tikzstyle{normalnode} = [dspnodefull,minimum size=1mm];
%\tikzstyle{reducednode} = [dspnodefull,minimum size=1mm];
\tikzstyle{normalline} = [line width = 0.5mm];
\tikzstyle{halfline} = [line width = 0.5mm, dashed];
\tikzstyle{conn} = [dspconn, line width = 0.5mm];

\node[frozennode] (u0) at (0.00, 7) {};
\node[left] at (0.00, 7) {$ u_{0} $};
\node[frozennode] (u1) at (0.00, 6) {};
\node[left] at (0.00, 6) {$ u_{1} $};
\node[frozennode] (u2) at (0.00, 5) {};
\node[left] at (0.00, 5) {$ u_{2} $};
\node[normalnode] (u3) at (0.00, 4) {};
\node[left] at (0.00, 4) {$ u_{3} $};
\node[frozennode] (u4) at (0.00, 3) {};
\node[left] at (0.00, 3) {$ u_{4} $};
\node[normalnode] (u5) at (0.00, 2) {};
\node[left] at (0.00, 2) {$ u_{5} $};
\node[normalnode] (u6) at (0.00, 1) {};
\node[left] at (0.00, 1) {$ u_{6} $};
\node[normalnode] (u7) at (0.00, 0) {};
\node[left] at (0.00, 0) {$ u_{7} $};
\node[dspadder] (node0) at (1.00, 7) {};
\node[normalnode] (node1) at (1.00, 6) {};
\draw[normalline] (u0)--(node0);
\draw[normalline] (u1)--(node1);
\draw[normalline] (node1)--(node0);
\node[dspadder] (node2) at (1.00, 5) {};
\node[normalnode] (node3) at (1.00, 4) {};
\draw[normalline] (u2)--(node2);
\draw[normalline] (u3)--(node3);
\draw[normalline] (node3)--(node2);
\node[dspadder] (node4) at (1.00, 3) {};
\node[normalnode] (node5) at (1.00, 2) {};
\draw[normalline] (u4)--(node4);
\draw[normalline] (u5)--(node5);
\draw[normalline] (node5)--(node4);
\node[dspadder] (node6) at (1.00, 1) {};
\node[normalnode] (node7) at (1.00, 0) {};
\draw[normalline] (u6)--(node6);
\draw[normalline] (u7)--(node7);
\draw[normalline] (node7)--(node6);
\node[dspadder] (node8) at (2.50, 7) {};
\node[normalnode] (node9) at (2.50, 5) {};
\draw[normalline] (node0)--(node8);
\draw[normalline] (node2)--(node9);
\draw[normalline] (node9)--(node8);
\node[dspadder] (node10) at (3.00, 6) {};
\node[normalnode] (node11) at (3.00, 4) {};
\draw[normalline] (node1)--(node10);
\draw[normalline] (node3)--(node11);
\draw[normalline] (node11)--(node10);
\node[dspadder] (node12) at (2.50, 3) {};
\node[normalnode] (node13) at (2.50, 1) {};
\draw[normalline] (node4)--(node12);
\draw[normalline] (node6)--(node13);
\draw[normalline] (node13)--(node12);
\node[dspadder] (node14) at (3.00, 2) {};
\node[normalnode] (node15) at (3.00, 0) {};
\draw[normalline] (node5)--(node14);
\draw[normalline] (node7)--(node15);
\draw[normalline] (node15)--(node14);
\node[dspadder] (node16) at (4.50, 7) {};
\node[normalnode] (node17) at (4.50, 3) {};
\draw[normalline] (node8)--(node16);
\draw[normalline] (node12)--(node17);
\draw[normalline] (node17)--(node16);
\node[dspadder] (node18) at (5.00, 6) {};
\node[normalnode] (node19) at (5.00, 2) {};
\draw[normalline] (node10)--(node18);
\draw[normalline] (node14)--(node19);
\draw[normalline] (node19)--(node18);
\node[dspadder] (node20) at (5.50, 5) {};
\node[normalnode] (node21) at (5.50, 1) {};
\draw[normalline] (node9)--(node20);
\draw[normalline] (node13)--(node21);
\draw[normalline] (node21)--(node20);
\node[dspadder] (node22) at (6.00, 4) {};
\node[normalnode] (node23) at (6.00, 0) {};
\draw[normalline] (node11)--(node22);
\draw[normalline] (node15)--(node23);
\draw[normalline] (node23)--(node22);
\node[normalnode] (x0) at (7.00, 7) {};
\node[right] at (7.00, 7) {$ x_{0} $};
\draw[normalline] (node16)--(x0);
\node[normalnode] (x1) at (7.00, 6) {};
\node[right] at (7.00, 6) {$ x_{1} $};
\draw[normalline] (node18)--(x1);
\node[normalnode] (x2) at (7.00, 5) {};
\node[right] at (7.00, 5) {$ x_{2} $};
\draw[normalline] (node20)--(x2);
\node[normalnode] (x3) at (7.00, 4) {};
\node[right] at (7.00, 4) {$ x_{3} $};
\draw[normalline] (node22)--(x3);
\node[normalnode] (x4) at (7.00, 3) {};
\node[right] at (7.00, 3) {$ x_{4} $};
\draw[normalline] (node17)--(x4);
\node[normalnode] (x5) at (7.00, 2) {};
\node[right] at (7.00, 2) {$ x_{5} $};
\draw[normalline] (node19)--(x5);
\node[normalnode] (x6) at (7.00, 1) {};
\node[right] at (7.00, 1) {$ x_{6} $};
\draw[normalline] (node21)--(x6);
\node[normalnode] (x7) at (7.00, 0) {};
\node[right] at (7.00, 0) {$ x_{7} $};
\draw[normalline] (node23)--(x7);

%\node[frozennode] (u0) at (9.00, 7) {};
%\node[left] at (9.00, 7) {$ u_{0} $};
%\node[frozennode] (u1) at (9.00, 6) {};
%\node[left] at (9.00, 6) {$ u_{1} $};
%\node[frozennode] (u2) at (9.00, 5) {};
%\node[left] at (9.00, 5) {$ u_{2} $};
\node[normalnode] (u3) at (9.00, 4) {};
\node[left] at (9.00, 4) {$ u_{3} $};
%\node[frozennode] (u4) at (9.00, 3) {};
%\node[left] at (9.00, 3) {$ u_{4} $};
\node[normalnode] (u5) at (9.00, 2) {};
\node[left] at (9.00, 2) {$ u_{5} $};
\node[normalnode] (u6) at (9.00, 1) {};
\node[left] at (9.00, 1) {$ u_{6} $};
\node[normalnode] (u7) at (9.00, 0) {};
\node[left] at (9.00, 0) {$ u_{7} $};
%\node[dspadder] (node0) at (10.00, 7) {};
%\node[normalnode] (node1) at (10.00, 6) {};
%\draw[normalline] (u0)--(node0);
%\draw[normalline] (u1)--(node1);
%\draw[normalline] (node1)--(node0);
\coordinate (node2) at (10.00, 5) {};
\node[normalnode] (node3) at (10.00, 4) {};
%\draw[normalline] (u2)--(node2);
\draw[halfline] (u3)--(node3);
\draw[normalline] (node3)--(node2);
\coordinate (node4) at (10.00, 3) {};
\node[normalnode] (node5) at (10.00, 2) {};
%\draw[normalline] (u4)--(node4);
\draw[halfline] (u5)--(node5);
\draw[normalline] (node5)--(node4);
\node[dspadder] (node6) at (10.00, 1) {};
\node[normalnode] (node7) at (10.00, 0) {};
\draw[halfline] (u6)--(node6);
\draw[halfline] (u7)--(node7);
\draw[halfline] (node7)--(node6);
\coordinate (node8) at (11.50, 7) {};
\node[normalnode] (node9) at (11.50, 5) {};
%\draw[normalline] (node0)--(node8);
\draw[normalline] (node2)--(node9);
\draw[normalline] (node9)--(node8);
\coordinate (node10) at (12.00, 6) {};
\node[normalnode] (node11) at (12.00, 4) {};
%\draw[normalline] (node1)--(node10);
\draw[normalline] (node3)--(node11);
\draw[normalline] (node11)--(node10);
\node[dspadder] (node12) at (11.50, 3) {};
\node[normalnode] (node13) at (11.50, 1) {};
\draw[normalline] (node4)--(node12);
\draw[halfline] (node6)--(node13);
\draw[normalline] (node13)--(node12);
\node[dspadder] (node14) at (12.00, 2) {};
\node[normalnode] (node15) at (12.00, 0) {};
\draw[normalline] (node5)--(node14);
\draw[halfline] (node7)--(node15);
\draw[normalline] (node15)--(node14);
\node[dspadder] (node16) at (13.50, 7) {};
\node[normalnode] (node17) at (13.50, 3) {};
\draw[normalline] (node8)--(node16);
\draw[normalline] (node12)--(node17);
\draw[normalline] (node17)--(node16);
\node[dspadder] (node18) at (14.00, 6) {};
\node[normalnode] (node19) at (14.00, 2) {};
\draw[normalline] (node10)--(node18);
\draw[normalline] (node14)--(node19);
\draw[normalline] (node19)--(node18);
\node[dspadder] (node20) at (14.50, 5) {};
\node[normalnode] (node21) at (14.50, 1) {};
\draw[normalline] (node9)--(node20);
\draw[normalline] (node13)--(node21);
\draw[normalline] (node21)--(node20);
\node[dspadder] (node22) at (15.00, 4) {};
\node[normalnode] (node23) at (15.00, 0) {};
\draw[normalline] (node11)--(node22);
\draw[normalline] (node15)--(node23);
\draw[normalline] (node23)--(node22);
\node[normalnode] (x0) at (16.00, 7) {};
\node[right] at (16.00, 7) {$ x_{0} $};
\draw[normalline] (node16)--(x0);
\node[normalnode] (x1) at (16.00, 6) {};
\node[right] at (16.00, 6) {$ x_{1} $};
\draw[normalline] (node18)--(x1);
\node[normalnode] (x2) at (16.00, 5) {};
\node[right] at (16.00, 5) {$ x_{2} $};
\draw[normalline] (node20)--(x2);
\node[normalnode] (x3) at (16.00, 4) {};
\node[right] at (16.00, 4) {$ x_{3} $};
\draw[normalline] (node22)--(x3);
\node[normalnode] (x4) at (16.00, 3) {};
\node[right] at (16.00, 3) {$ x_{4} $};
\draw[normalline] (node17)--(x4);
\node[normalnode] (x5) at (16.00, 2) {};
\node[right] at (16.00, 2) {$ x_{5} $};
\draw[normalline] (node19)--(x5);
\node[normalnode] (x6) at (16.00, 1) {};
\node[right] at (16.00, 1) {$ x_{6} $};
\draw[normalline] (node21)--(x6);
\node[normalnode] (x7) at (16.00, 0) {};
\node[right] at (16.00, 0) {$ x_{7} $};
\draw[normalline] (node23)--(x7);

\end{tikzpicture}}
	\vspace{-0.5cm}
	\caption{\footnotesize Forney-style factor graph (FFG) of the RM(1,3)-code (left) and the reduced FFG (right). Note that the dashed lines indicate variables to be computed only in the right-to-left message update.}
	\label{fig:rm_ffg}
	\vspace{-0.5cm}	
\end{figure}

\subsubsection{Minimum Weight Parity-Check Belief Propagation Decoding}
Minimum weight parity-check belief propagation\acused{MWPC-BP} (\ac{MWPC-BP}) decoding introduced in \cite{PfisterMWPC} is based on the concept of iterative decoding over an overcomplete parity-check matrix. An online algorithm tailored to the noisy received sequence $ \yv $ is used to construct the overcomplete parity-check matrix only based on minimum weight parity-checks.
\subsubsection{Neural Belief Propagation Decoding}
Neural belief propagation\acused{NBP} (\ac{NBP}) decoding as introduced in \cite{NachmaniNeuralBP} treats the unrolled Tanner graph of the code as a neural network (NN), while assigning \emph{trainable} weights to all of its edges leading to a \emph{soft} Tanner graph. These trainable weights are optimized via \ac{SGD} methods.
\subsubsection{Pruned Neural Belief Propagation Decoding}
Pruned neural belief propagation (pruned-\ac{NBP}) decoding \cite{buchberger2020pruning} combines the idea of \ac{MWPC-BP} together with \ac{NBP}. To get started, a redundant parity-check matrix containing (all or some of) the minimum weight parity-checks is constructed. During the offline training phase, all edges connected to a check node are assigned a single trainable weight and the least contributing check nodes are \emph{pruned} (i.e., removed) from the graph. The authors of \cite{buchberger2020pruning} refer to this decoder as $ D_1 $. An enhanced version,  decoder $ D_3 $, is the result of assigning trainable weights per edge \new{at} the expense of larger memory requirements to save all weights per edges. Furthermore, a pruned \ac{NBP} decoder without any weights is introduced as $ D_2 $, however with the expense of a significant degradation in error-rate performance.

\section{Automorphism Ensemble Decoding}\label{sec:ensemble_dec}
Ensemble decoding uses multiple constituent decoders (i.e., a decoder \emph{ensemble} of size $ M $) to generate a set of codeword estimates and selects one of these codewords, using a predefined metric, as the decoder output. Typically, a least-squares metric is used, as this corresponds to the \ac{ML} decision for the \ac{AWGN} channel. Hence, this method is also called \emph{ML-in-the-list}. This can be formulated as 
\begin{equation} \label{eq:mlinthelist}
\hat{\xv} = \argmin_{\hat{\xv}_j, j \in \LP 1, \dots, M\RP} \left\Vert \hat{\xv}_j - \yv \right\Vert^2 = \argmax_{\hat{\xv}_j, j \in \LP 1, \dots, M\RP} \sum_{i=0}^{N-1} \hat{x}_{j,i} \cdot y_i,
\end{equation} % keep this to explain the complexity
where $ \hat{x}_{j,i} \in \LP \pm 1 \RP$, $\hat{\xv}_j$ is the estimated codeword from decoder~$j$ for the received vector $ \yv $ and $\hat{\xv}$ is the final codeword estimate of the ensemble.

\ac{MBBP} is a well-known example for ensemble decoding that uses $ M $ \ac{BP} decoders, each based on a different random parity-check matrix \cite{Huber}. Another instance of ensemble decoding is \ac{BPL} decoding of polar codes, where the stages of the \ac{FFG} are randomly permuted for each constituent decoder \cite{elkelesh2018belief}. 

In this work, we propose \emph{automorphism ensemble BP decoding} (Aut-$M$-BP) for RM codes. The main idea is to make use of the already existent polar \ac{BP} decoder. Furthermore, we use the RM code symmetry in the decoding algorithm itself, as permuting a valid RM codeword with a permutation from the code's automorphism group results in another valid RM codeword.

An abstract view of our proposed Aut-$M$-BP decoding algorithm is shown in Fig.~\ref{fig:Block_Diag}.
The input to the decoder is the received noisy codeword $\yv$. 
We randomly sample $M$ different permutations from the RM automorphism group, where each permutation is denoted by $\pi_j$, with $j$ being the decoder index and $j \in \{1,2,\cdots,M\}$.
The $\yv$-vector is interleaved (i.e., permuted) with the $M$ different permutations $\pi_j$ leading to $M$ permuted noisy codewords $\yv'_j$, where $j \in \{1,2,\cdots,M\}$. 
Now we decode every $\yv'_j$-vector independently using \ac{BP}, whose output is the interleaved estimated codeword $\hat{\xv}'_j$.
A de-interleaving phase is applied to all $M$ interleaved estimated codewords $\hat{\xv}'_j$ and, thus, we have $M$ codeword estimates $\hat{\xv}_j$. Let $ \operatorname{BP}(\cdot) $ denote the \ac{BP} decoding function that maps $ \yv'_j $ to $ \hat{\xv}_j' $. Then we can write the interleaved decoding as
\begin{equation}\label{eq:aut_dec}
\hat{\xv}_j = \pi^{-1}_j\left(\operatorname{BP}\left(\pi_j(\yv)\right)\right).
\end{equation}

Similar to \ac{MBBP} and \ac{BPL} decoding, our proposed decoding algorithm uses the \emph{ML-in-the-list} picking rule according to Eq.~(\ref{eq:mlinthelist}) to choose the most likely codeword from the list to get the final decoder output $\hat{\xv}$.

As the decoders are linear, their decoding behavior is only dependent on the noise induced by the channel, and not the choice of the transmitted codeword. Therefore, decoding using automorphisms according to Eq.~(\ref{eq:aut_dec}) corresponds to permuting the \emph{noise}. It is reasonable to conclude that suboptimal (i.e., not ML) decoders may react differently to noise realizations in different permutations, which is exactly the property that automorphism ensemble decoding seeks to exploit.

Our proposed algorithm can be therefore seen as a natural generalization of the \ac{BPL} decoder \cite{elkelesh2018belief}: We still use $M$ parallel independent BP decoders. However, we use a more general set of permutations. It was shown in \cite{Doan_2018_Permuted_BP} that the stage-shuffling of the \ac{FFG} is equivalent to a bit-interleaving operation while keeping the factor graph unchanged; these permutations correspond to the automorphism subgroup $ \Pi(m) $. In contrast, we use permutations from the whole RM code automorphism group $ \operatorname{GA}(m) $ (rather than only $\Pi(m)$, which is used in \ac{BPL} decoding as proposed in \cite{elkelesh2018belief}). 

It is worth mentioning that the usage of a \ac{BP} decoder as a constituent decoder has some similarities when compared to \emph{automorphism group decoding} of \ac{BCH} and Golay codes for the BEC \cite{HUBER_BCH_Golay} and for the AWGN channel \cite{Dimnik_RRD_HDPC}.
Automorphism group decoding is based on permuting the received sequence exploiting automorphisms of the code while applying an iterative message passing algorithm. 

\new{\begin{figure}[t]
		\begin{tikzpicture}
\begin{axis}[
	width=\linewidth,
	height=.9\linewidth,
	grid style={dotted,anthrazit},
	xmajorgrids,
	yminorticks=true,
	ymajorgrids,
	%yminorgrids,
	legend columns=1,
	%legend pos=south west,   
	legend cell align={left},
	legend style={draw=none, fill=none, row sep=.4mm, at={(0.03,-.01)},anchor=south west}, %fill,fill opacity=0.8},
	xlabel={$E_\mathrm{b}/N_0$ [dB]},
	ylabel={BLER},
	legend image post style={mark indices={}},% <- added	
	ymode=log,
	mark size=1.5pt,
	xmin=2,
	xmax=4.5,
	ymin=3e-8,
	ymax=9.756e-01
]

\addplot[color=pink,line width = 1pt, dotted, mark size=2.5pt, mark options={solid}]
table[col sep=comma]{
1.00, 9.756e-01
1.50, 9.221e-01
2.00, 8.242e-01
2.50, 6.682e-01
3.00, 4.848e-01
3.50, 3.109e-01
4.00, 1.801e-01
4.50, 9.100e-02
5.00, 3.890e-02
5.50, 1.750e-02
};
\label{plot:rm_bp_naive}
\addlegendentry{\footnotesize Na\"ive BP };
\addplot[color=mittelblau,line width = 1pt, dotted,mark=square,mark size=2.5pt, mark options={solid}]
table[col sep=comma]{
1.00, 6.909e-01
1.50, 4.966e-01
2.00, 2.907e-01
2.50, 1.211e-01
3.00, 3.608e-02
3.50, 9.485e-03
4.00, 2.122e-03
4.50, 5.311e-04
5.00, 1.372e-04
};
\label{plot:rm_bp}
\addlegendentry{\footnotesize FFG BP};

\addplot[color=black,line width = 1pt, dashed,mark size=2.5pt, mark options={solid}]
table[col sep=comma]{
	2.51, 1.612e-02
	3.01, 7.184e-03
	3.51, 2.825e-03
	4.01, 7.922e-04
	4.51, 1.828e-04
	4.71, 1.045e-04
};
\label{plot:rm_mbbp}
\addlegendentry{\footnotesize MBBP-60 \cite{buchberger2020pruning}};

\addplot [color=blue,line width = 1pt, dashed, mark=+, mark size=2.5pt, mark options={solid}]
table[row sep=crcr]{%
	2	0.0512291666666667\\
	2.5	0.011875\\
	3	0.00207945736434109\\
	3.5	0.000313669064748201\\
	4	3.7099161322151e-05\\
	4.5	2.69867196932036e-06\\
};
\addlegendentry{\footnotesize MWPC-BP \cite{PfisterMWPC}}

\addplot[color=apfelgruen,line width = 1pt, dashed,mark=x,mark size=2.5pt, mark options={solid}]
table[col sep=comma]{
	1.00, 4.807e-01
	1.50, 2.611e-01
	2.00, 9.246e-02
	2.50, 2.340e-02
	3.00, 4.018e-03
	3.50, 5.299e-04
	4.00, 4.559e-05
	4.50, 3.912e-06
};
\label{plot:rm_bpl8}
\addlegendentry{\footnotesize BPL-8 \ref{plot:rm_bpl32} BPL-32 };

\addplot[color=apfelgruen,line width = 1pt,mark=x, solid,mark size=2.5pt, mark options={solid}, forget plot]
table[col sep=comma]{
	1.00, 3.809e-01
	1.50, 1.679e-01
	2.00, 5.829e-02
	2.50, 1.427e-02
	3.00, 1.985e-03
	3.50, 2.119e-04
	4.00, 1.863e-05
	4.50, 1.579e-06
};
\label{plot:rm_bpl32}
%\addlegendentry{\footnotesize BPL-32};

\addplot[color=mittelblau,line width = 1pt, dashed,mark=o,mark size=2.5pt, mark options={solid}]
table[col sep=comma]{
	1.00, 3.599e-01
	1.50, 1.766e-01
	2.00, 6.498e-02
	2.50, 1.325e-02
	3.00, 2.103e-03
	3.50, 1.946e-04
	4.00, 1.452e-05
	4.50, 9.887e-07
};
\label{plot:rm_aut_8_bp}
\addlegendentry{\footnotesize Aut-8-BP \ref{plot:rm_aut_32_bp} Aut-32-BP };

\addplot[color=mittelblau,line width = 1pt,mark=o, solid,mark size=2.5pt, mark options={solid}, forget plot]
table[col sep=comma]{
	1.00, 2.776e-01
	1.50, 1.076e-01
	2.00, 3.173e-02
	2.50, 6.082e-03
	3.00, 1.039e-03
	3.50, 1.247e-04
	4.00, 1.139e-05
	4.50, 9.293e-07
};
\label{plot:rm_aut_32_bp}
%\addlegendentry{\footnotesize Aut-32-BP};

\addplot[color=gray,line width = 1pt, solid,mark size=2.5pt, mark options={solid}]
table[col sep=comma, row sep=crcr]{
	1.00, 1.880e-01\\
	1.50, 7.969e-02\\
	2.00, 2.533e-02\\
	2.50, 6.105e-03\\
	3.00, 1.105e-03\\
	3.50, 1.333e-04\\
	4.00, 1.275e-05\\
};
\label{plot:rm_scl32}
\addlegendentry{\footnotesize SCL-32};

\addplot[color=black,line width = 0.8pt, solid,mark size=2.5pt,mark=star, mark options={solid}]
table[col sep=comma]{
	1.00, 1.900e-01
	1.50, 7.670e-02
	2.00, 2.276e-02
	2.50, 5.760e-03
	3.00, 1.026e-03
	3.50, 1.324e-04
	4.00, 1.450e-05
};
\label{plot:rm_ds32}
\addlegendentry{\footnotesize \new{DS-32 ($ \Pi $) \cite{permuteRM}} };

\addplot[color=rot,line width = 1pt,mark=x, dashed,mark size=2.5pt, mark options={solid}]
table[col sep=comma]{
	0.00, 5.025e-01
	0.50, 3.205e-01
	1.00, 1.538e-01
	1.50, 5.590e-02
	2.00, 1.538e-02
	2.50, 4.031e-03
	3.00, 6.489e-04
	3.50, 9.700e-05
};
\label{plot:rm_ml}
\addlegendentry{\footnotesize ML \cite{kldatabase} \ref{plot:rm_osd4} OSD-4};

\addplot[color=rot, line width = 1pt, solid, mark size=2.5pt, mark options={solid}]
table[col sep=comma]{
3.50, 1.063e-04
4.00, 1.090e-05
};
\label{plot:rm_osd4}
%\addlegendentry{\footnotesize OSD-4};

\end{axis}

\end{tikzpicture}
		\vspace{-0.7cm}
		\caption{\footnotesize BLER comparison between non \ac{SGD}-optimized iterative decoders, recursive list decoding and our proposed Aut-BP decoding scheme for the RM(3,7)-code over the BI-AWGN channel.}  
		\label{fig:RM_BLER_classic}
		\vspace{-0.2cm}
\end{figure}
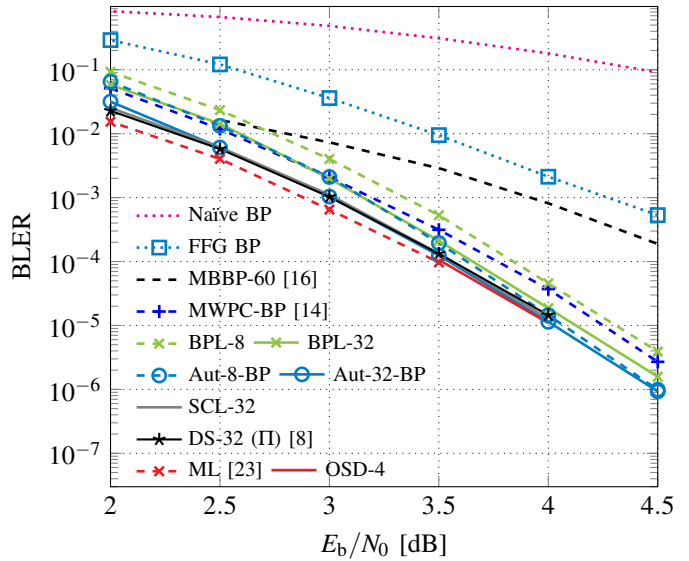}
\section{Results}\label{sec:Results}
Regarding practical applications, both error-rate performance and the computational complexity of the decoding scheme have to be considered. We compare the described decoding schemes for the RM(3,7)-code with $ N=128 $ and $ k=64 $. In the following, we specify the parameters of some of the compared decoders for reproducibility:
\begin{itemize}
\item \textbf{\ac{MWPC-BP}} utilizes 5\% of the minimum-weight parity-checks, as reported in \cite{PfisterMWPC}.
\item \textbf{\ac{MBBP}} operates over $ M=60 $ randomly generated parity-check matrices with 6 iterations each. 
\item \textbf{Neural-BP} uses all 94488 minimum-weight parity-checks over 6 iterations. 
\item The \textbf{pruned neural-BP} employs on average 3\% of the minimum-weight parity-checks over a total of 6 iterations. We consider the three variants of this decoder as introduced in \cite{buchberger2020pruning}, with tied weights ($ D_1 $), no weights ($ D_2 $) and free weights ($ D_3 $). 
\item For our proposed \textbf{Aut-BP}, we show results for both $ M=8 $ and $ M=32 $ randomly chosen permutations from the full automorphism group. Here, up to $ N_\mathrm{it,max} = 200 $ iterations are performed with, however, an early stopping condition (i.e., when $ \hat{\xv}=\hat{\uv}\Gm $) employed to reduce the average total number of iterations. Furthermore, the \acp{FFG} have been reduced from 1792 to 1334 box-plus and addition operations per full iteration by removing operations with constant results, as presented in Section \ref{sec:ffg}. 
\end{itemize}

\subsection{Error-Rate Performance}
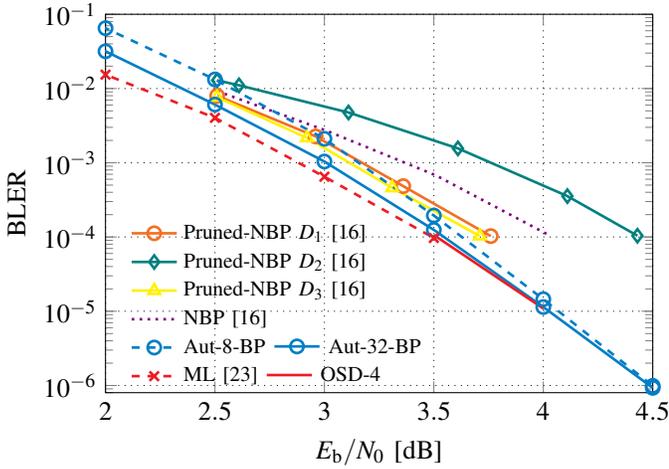
\begin{figure}[t]
	\begin{tikzpicture}
\begin{axis}[
	width=\linewidth,
	height=.747\linewidth,
	grid style={dotted,anthrazit},
	xmajorgrids,
	yminorticks=true,
	ymajorgrids,
	%yminorgrids,
	legend columns=1,
	%legend pos=south west,   
	legend cell align={left},
	legend style={draw=none, fill=none, row sep=-.05mm, at={(0.03,-.01)},anchor=south west}, 
	xlabel={$E_\mathrm{b}/N_0$ [dB]},
	ylabel={BLER},
	legend image post style={mark indices={}},% <- added	
	ymode=log,
	mark size=1.5pt,
	xmin=2,
	xmax=4.5,
	ymin=8e-7,
	ymax=1e-1
]

\addplot[color=orange,line width = 1pt, mark=o, solid,mark size=2.5pt, mark options={solid}]
table[col sep=comma]{
2.51, 7.978e-03
2.96, 2.263e-03
3.36, 4.846e-04
3.76, 1.028e-04
};
\label{plot:rm_d1}
\addlegendentry{\footnotesize Pruned-NBP $ D_1 $ \cite{buchberger2020pruning}};

\addplot[color=teal,line width = 1pt, mark=diamond, solid,mark size=2.5pt, mark options={solid}]
table[col sep=comma]{
2.51, 1.289e-02
2.61, 1.104e-02
3.11, 4.756e-03
3.61, 1.562e-03
4.11, 3.554e-04
4.43, 1.041e-04
};
\label{plot:rm_d2}
\addlegendentry{\footnotesize Pruned-NBP $ D_2 $ \cite{buchberger2020pruning}};

\addplot[color=yellow,line width = 1pt, mark=triangle, solid,mark size=2.5pt, mark options={solid}]
table[col sep=comma]{
2.51, 7.642e-03
2.92, 2.164e-03
3.31, 4.597e-04
3.71, 1.030e-04
};
\label{plot:rm_d3}
\addlegendentry{\footnotesize Pruned-NBP $ D_3 $ \cite{buchberger2020pruning}};

\addplot[color=violet,line width = 1pt, dotted,mark size=2.5pt, mark options={solid}]
table[col sep=comma]{
2.51, 9.236e-03
3.01, 2.698e-03
3.51, 6.749e-04
4.03, 1.044e-04

};
\label{plot:rm_wbp}
\addlegendentry{\footnotesize NBP \cite{buchberger2020pruning}};

\addplot[color=mittelblau,line width = 1pt, dashed,mark=o,mark size=2.5pt, mark options={solid}]
table[col sep=comma]{
	1.00, 3.599e-01
	1.50, 1.766e-01
	2.00, 6.498e-02
	2.50, 1.325e-02
	3.00, 2.103e-03
	3.50, 1.946e-04
	4.00, 1.452e-05
	4.50, 9.887e-07
};
\label{plot:rm_aut_8_bp}
\addlegendentry{\footnotesize Aut-8-BP \ref{plot:rm_aut_32_bp} Aut-32-BP };

\addplot[color=mittelblau,line width = 1pt,mark=o, solid,mark size=2.5pt, mark options={solid}, forget plot]
table[col sep=comma]{
	1.00, 2.776e-01
	1.50, 1.076e-01
	2.00, 3.173e-02
	2.50, 6.082e-03
	3.00, 1.039e-03
	3.50, 1.247e-04
	4.00, 1.139e-05
	4.50, 9.293e-07
};
\label{plot:rm_aut_32_bp}
%\addlegendentry{\footnotesize Aut-32-BP};

\addplot[color=rot,line width = 1pt,mark=x, dashed,mark size=2.5pt, mark options={solid}]
table[col sep=comma]{
	0.00, 5.025e-01
	0.50, 3.205e-01
	1.00, 1.538e-01
	1.50, 5.590e-02
	2.00, 1.538e-02
	2.50, 4.031e-03
	3.00, 6.489e-04
	3.50, 9.700e-05
};
\label{plot:rm_ml}
\addlegendentry{\footnotesize ML \cite{kldatabase} \ref{plot:rm_osd4} OSD-4};

\addplot[color=rot, line width = 1pt, solid, mark size=2.5pt, mark options={solid}]
table[col sep=comma]{
	3.50, 1.063e-04
	4.00, 1.090e-05
};
\label{plot:rm_osd4}
%\addlegendentry{\footnotesize OSD-4};

\end{axis}

\end{tikzpicture}
	\vspace{-0.7cm}
	\caption{\footnotesize BLER comparison between \ac{SGD}-optimized (NN-based) iterative decoders and Aut-BP for the RM(3,7)-code over the BI-AWGN channel. All neural-\ac{BP} decoders use $ N_\mathrm{it} = 6 $ iterations.} 
	\label{fig:RM_BLER_neural}
	\vspace{-0.2cm}
\end{figure}

In Fig.~\ref{fig:RM_BLER_classic} and Fig. \ref{fig:RM_BLER_neural},  we showcase the error-rate performance of the described decoding schemes for the RM(3,7)-code over the \ac{AWGN} channel using \ac{BPSK} mapping. Furthermore, we show the \ac{ML} performance of the code as provided by \cite{kldatabase}. As no data beyond an \ac{SNR} of 3.5 dB is available, the \ac{ML} performance is estimated using order-4 \ac{OSD}. 

Fig.~\ref{fig:RM_BLER_classic} compares the non-\ac{SGD}-optimized iterative decoders with Aut-BP and \ac{ML}. One can observe that the na\"ive \ac{BP} decoding suffers from a very poor performance for \ac{RM} codes, compared to \ac{BP} decoding over \ac{FFG}. Moreover, using multiple $ \Hm $-matrices in \ac{MBBP} leads to a significant enhancement in performance. All of the previous methods are outperformed by both Aut-8-BP and \ac{MWPC-BP}, with similar performance. However, in the high \ac{SNR} regime, Aut-8-BP beats \ac{MWPC-BP} by 0.2~dB. Aut-32-BP even closes the gap to \ac{ML} to less than 0.05~dB at a \ac{BLER} of $ 10^{-4} $.
We further observe the gains of sampling from $ \operatorname{GA}(m) $ in Aut-BP compared to $ \Pi(m) $, as used in \ac{BPL} decoding. We can see that for all ensemble sizes $ M $, sampling from $ \operatorname{GA}(m) $ consistently outperforms $ \Pi(m) $ by up to 0.3 dB. This confirms the sub-optimality of restricting the automorphisms to a small subgroup. Moreover, Aut-32-BP can even outperform SCL \cite{talvardyList} with list size $L=32$ (i.e., SCL-32) and its permutation variant DS-32 \cite{permuteRM} in the high SNR regime.

Fig.~\ref{fig:RM_BLER_neural} compares the \ac{SGD}-optimized (NN-based) decoders with Aut-BP and \ac{ML}. Here, the neural-BP decoder is much closer to the \ac{ML} bound, and the pruned variant with free weights $ D_3 $ outperforms \ac{NBP}, which uses all overcomplete parity-checks. The pruned \ac{NBP} $ D_2 $ decoder without weights suffers from a significant performance degradation. Over the whole SNR range, $ D_1 $ and $ D_3 $ are outperformed by Aut-32-BP. Furthermore, it can be seen that using only $ M=8 $ parallel BP decoders (i.e., Aut-8-BP) results in a small performance degradation of less than 0.2~dB over the whole \ac{SNR} range, offering an attractive trade-off for lower complexity. Simulation results for the RM(4,8)-code are presented in \cite{rm_automorphism_ensemble_decoding}.

\subsection{Iterative Decoding Complexity}\label{ssec:Complexity}
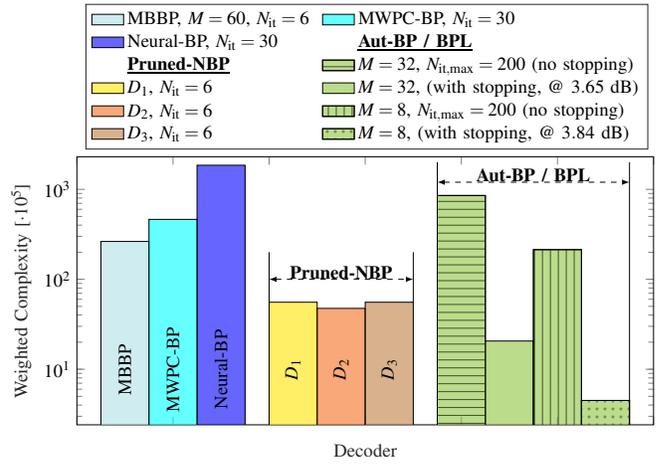
\begin{figure}[t]
	\resizebox{0.48\textwidth}{!}{\definecolor{mycolor1}{rgb}{0.00000,0.44700,0.74100}%
\definecolor{mycolor2}{rgb}{0.85000,0.32500,0.09800}%
\definecolor{mycolor3}{rgb}{0.69, 0.878,0.902}%
\begin{tikzpicture}

\begin{axis}[%
width=4.3in,
height=2.0in,
scale only axis,
xmin=0,
xmax=12,
ymin=0,
ymax=2300,
ymode=log,
%     yminorticks=true,
%     ymajorgrids,
axis background/.style={fill=white},
xlabel style={font=\color{white!15!black}},
xlabel={Decoder},
xticklabels={},
ylabel style={font=\color{white!15!black}},
ylabel={Weighted Complexity [$ \cdot 10^5 $]},
legend columns=6, 
transpose legend,
legend style={at={(0.015,1.57)}, anchor=north west, legend cell align=left, align=left, draw=white!15!black,/tikz/column 2/.style={column sep=5pt}}
]

\addplot[ybar interval, fill=mycolor3, fill opacity=0.6, draw=black, area legend] table[row sep=crcr] {%
	x	y\\
	0.5   263.4\\
	1.5   263.4\\	
};
\addlegendentry{MBBP, $M=60$, $ N_\mathrm{it}=6 $}
\node[rotate=90] at (axis cs: 1,10) {MBBP};

\addplot[ybar interval, fill=blue, fill opacity=0.6, draw=black, area legend] table[row sep=crcr] {%
	x	y\\
	2.5   1853.9\\
	3.5   1853.9\\	
};
\addlegendentry{Neural-BP, $ N_\mathrm{it}=30 $ }
\node[rotate=90] at (axis cs: 3,10) {Neural-BP};

\addlegendimage{empty legend}
\addlegendentry{\hspace{0cm}\textbf{\underline{Pruned-NBP}}}

\addplot[ybar interval, fill=gelb, fill opacity=0.6, draw=black, area legend] table[row sep=crcr] {%
	x	y\\
	4   55.6\\
	5   55.6\\	
};
\addlegendentry{$ D_1 $, $ N_\mathrm{it}=6 $}
\node[rotate=90] at (axis cs: 4.5,10) {$ D_1 $};

\addplot[ybar interval, fill=orange, fill opacity=0.6, draw=black, area legend] table[row sep=crcr] {%
	x	y\\
	5   47.5\\
	6   47.5\\	
};
\addlegendentry{$ D_2 $, $ N_\mathrm{it}=6 $}
\node[rotate=90] at (axis cs: 5.5,10) {$ D_2 $};

\addplot[ybar interval, fill=brown, fill opacity=0.6, draw=black, area legend] table[row sep=crcr] {%
	x	y\\
	6   55.6\\
	7   55.6\\	
};
\addlegendentry{$ D_3 $, $ N_\mathrm{it}=6 $}
\node[rotate=90] at (axis cs: 6.5,10) {$ D_3 $};

\node[] at (axis cs: 5.5,120) {\textbf{Pruned-NBP}};
\addplot[<->,black,line legend,sharp plot,update limits=false,dashed,forget plot] coordinates {(4,100) (7,100)};
\addplot[black,line legend,sharp plot,update limits=false,forget plot] coordinates {(4,55.6) (4,200)};
\addplot[black,line legend,sharp plot,update limits=false,forget plot] coordinates {(7,55.6) (7,200)};

\addplot[ybar interval, fill=aqua, fill opacity=0.6, draw=black, area legend] table[row sep=crcr] {%
	x	y\\
	1.5   463.6\\
	2.5   463.6\\
};
\addlegendentry{MWPC-BP, $ N_\mathrm{it}=30 $}
\node[rotate=90] at (axis cs: 2,10) {MWPC-BP};

\addlegendimage{empty legend}
\addlegendentry{\hspace{0cm}\textbf{\underline{Aut-BP / BPL}}} 

\addplot[ybar interval, fill=apfelgruen, fill opacity=0.6, draw=black, area legend, postaction={pattern=horizontal lines}] table[row sep=crcr] {%
	x	y\\
	7.5   853.8\\
	8.5   853.8\\	
};
\addlegendentry{$M=32$, $ N_\mathrm{it,max}=200 $ (no stopping)}

\addplot[ybar interval, fill=apfelgruen, fill opacity=0.6, draw=black, area legend] table[row sep=crcr] {%
	x	y\\
	8.5   20.6\\
	9.5   20.6\\	
};
\addlegendentry{$M=32$, (with stopping, @ $3.65$ dB)}

\addplot[ybar interval, fill=apfelgruen, fill opacity=0.6, draw=black, area legend, postaction={pattern=vertical lines}] table[row sep=crcr] {%
	x	y\\
	9.5   213.5\\
	10.5   213.5\\	
};
\addlegendentry{$M=8$, $ N_\mathrm{it,max}=200 $ (no stopping)}

\addplot[ybar interval, fill=apfelgruen, fill opacity=0.6, draw=black, area legend, postaction={pattern=dots}] table[row sep=crcr] {%
	x	y\\
	10.5  4.5\\
	11.5  4.5\\	
};
\addlegendentry{$M=8$, (with stopping, @ $3.84$ dB)}

\node[] at (axis cs: 9.5,1450) {\textbf{Aut-BP / BPL}};
\addplot[<->,black,line legend,sharp plot,update limits=false,dashed,forget plot] coordinates {(7.5,1200) (11.5,1200)};
\addplot[black,line legend,sharp plot,update limits=false,forget plot] coordinates {(7.5,853.8) (7.5,2000)};
\addplot[black,line legend,sharp plot,update limits=false,forget plot] coordinates {(11.5,4.5) (11.5,2000)};

\end{axis}
\end{tikzpicture}%
}
	\caption{\footnotesize Complexity comparison of different iterative decoders using basic operations weighted according to Table~\ref{tab:basic_ops} (e.g., weight for multiplication = 3) to reach a target BLER of $ 10^{-4} $; RM(3,7)-code; BI-AWGN channel.}
	\label{fig:complexity}
	\vspace{-2mm}
\end{figure}

For the RM(3,7)-code, we compare the complexity of the iterative decoding algorithms with error-rate performance close to \ac{ML} by counting the number of computing operations required to decode one \ac{RM} codeword. The first column of Table \ref{tab:basic_ops} gives the list of the operations we use.\footnote{Note that these operations differ from the ETSI basic operations, as we are more interested in hardware than in software implementations.} Furthermore, as non-trivial multiplication is significantly more complex than the other considered operations, we introduce a weighting factor for multiplication $ w_\mathrm{mul}=3 $ (equivalent to the number of full-adders in a 4+1 bit fixed-point implementation) to make the comparison more fair. For all decoders, we assume that the box-plus operation is implemented as
\begin{align}
L_1 \boxplus L_2 &= \sgn(L_1) \cdot \sgn(L_2) \cdot \min(|L_1|,|L_2|) \nonumber  \\
&\quad+ f_{+}(|L_1+L_2|) - f_{+}(|L_1-L_2|),
\end{align}
where $ f_{+}(|x|) = \log\LB1+\exp(-|x|)\RB$ is a correction term that can be well-approximated by a short \ac{LUT}. Furthermore, \acp{CN} are assumed to be efficiently implemented using the box-minus operator as
\begin{equation}
L_{j\to i} = \bigboxplus_{i' \neq i} L_{i'\to j} = \LB \bigboxplus_{i'} L_{i'\to j} \RB \boxminus L_{i\to j},
\end{equation}
with $ L_1 \boxminus L_2 = \sgn(L_2) \cdot L_1 + f_{-}(|L_1+L_2|) - f_{-}(|L_1-L_2|) $ and $ f_{-}(|x|) = \log(1-\exp(-|x|)) $ which is again implemented as a \ac{LUT} as proposed in \cite{VaryBoxMinus}. The remaining columns of Table \ref{tab:basic_ops} list the number of operations of each type required for the basic building blocks of the described iterative decoding algorithms, namely box-plus evaluations, \ac{CN} and \ac{VN} updates, the ML-in-the-list decision and the stopping condition that is used in \ac{BPL} decoding. Neural-\ac{BP}, the pruned neural-\ac{BP} and \ac{MWPC-BP} decoding use non-trivial multiplications with the corresponding weights before the \ac{VN} evaluations.

\begin{table*}[t]
	\caption{\footnotesize Basic operations and their usage in iterative decoding. \textsuperscript{a}For BP with weights (MWPC-BP, NBP, D1, D3). } \label{tab:basic_ops}
	\centering{\begin{tabular}{l|c|ccccc}
			Operation & Weight & 2-input $ \boxplus $ & CN (deg. $ D $) & VN (deg. $ D $) & ML out of $ M $ & FFG BP Stopping \\
			\hline
			$ \sgn(x)\cdot\sgn(y)$ & 1 & 1 & $ D-1 $ & 0 & 0 & $ m \cdot N/2 + 2N-1 $\\
			$ \sgn(x)\cdot y$ & 1 & 1 & $ 2D-1 $ & 0 & $ MN $ & 0\\
			$ \min(|x|,|y|) $ & 1 & 1 & $ D-1 $ & 0 & 0 & 0\\
			$ \max(x,y) $ & 1 & 0 & 0 & 0 & $ M-1 $ & 0\\
			$ f_{\pm}(|x|) $ (LUT) & 1 & 2 & $ 4D-2 $ & 0 & 0 & 0\\
			$ x+y $,  $ x-y $ & 1 & 4 & $ 8D-4 $ & $ 2D $ & $ M(N-1) $ & 0\\
			$ x\cdot y $ & 3 & 0 & 0 & [$ D+1 $]\textsuperscript{a}  & 0 & 0\\
			\hline
			Weighted total & - & 9 & $ 16D-9 $ & $ 2D $  [$ +3D+3 $]\textsuperscript{a}  & $ 2MN-1 $ & $ m \cdot N/2 + 2N-1 $
	\end{tabular}}
\end{table*}
 
Fig. \ref{fig:complexity} shows the total number of weighted operations to decode one codeword of the RM(3,7)-code. We can see that out of all methods, neural-BP using the full overcomplete $ \Hm $-matrix has the highest complexity. The corresponding pruned decoders $ D_1 $ and $ D_3 $ result in approximately 3\% of that complexity. \ac{MWPC-BP} is computationally more expensive, as it uses more parity-check equations and more iterations are required to achieve a good error-rate performance. It has to be noted, however, that we only list the complexity of iterative decoding, not of (adaptively) obtaining the parity-check equations. Hence, the overall complexity of \ac{MWPC-BP} is higher than the presented number. \ac{MBBP} has roughly half the complexity of \ac{MWPC-BP}, while Aut-32-BP without stopping condition has twice the complexity of \ac{MWPC-BP}. However, when a ($ \Gm $-matrix-based) stopping condition is used, the average number of iterations of Aut-BP is significantly reduced. To illustrate this, we measure the average required number of iterations until convergence for both $ M=8 $ and $ M=32 $ with respect to the \ac{SNR} of the \ac{AWGN} channel, while $ N_\mathrm{it,max} = 200$. At an \ac{SNR} of 3.65~dB, corresponding to the \ac{BLER} of $ 10^{-4} $, each decoder of the Aut-32-BP ensemble requires an average of 4.55 iterations, making Aut-BP the least complex decoder of the compared algorithms (see Fig.~\ref{fig:complexity}), without losing any error-rate performance. Aut-8-BP requires an \ac{SNR} of 3.84~dB to reach this \ac{BLER} performance, however, reducing the complexity again by a factor of 4, using only 3.96 iterations on average. Note that even though the ML-in-the-list decision can be only made after all constituent decoders are terminated, terminated decoders can already start decoding the next received vector (e.g., in a super-scalar implementation). 

It is also worth noting here that the approaches proposed in \cite{buchberger2020pruning} can also profit from early stopping. However, the effect is less significant since the majority of check node evaluations are performed in the first two iterations.
Moreover, the regular structure of the \ac{FFG} may result in more preferable implementations of Aut-BP compared to the random memory access patterns observed in conventional iterative decoders.
	
\section{Conclusion}\label{sec:conc}

In this work, we propose an automorphism-based iterative decoding algorithm for RM codes.
We present near-ML error-rate performance for the RM(3,7)-code operating only $0.05$ dB away from the ML bound at BLER of $10^{-4}$.
Furthermore, we report a decoder complexity comparison for the RM(3,7)-code from an operation level perspective.
To the best of our knowledge, our proposed iterative Aut-BP decoders using the RM code automorphism group as permutations are the best iterative decoders reported in literature thus far in terms of error-rate performance.

\bibliographystyle{IEEEtran}
\bibliography{references}
\end{NoHyper}
\end{document}